\documentclass[twocolumn,prb,showpacs]{revtex4-1}
\usepackage{times,xspace}
\usepackage{graphicx,graphics,color,epsfig}
\usepackage{bm}
\usepackage{amsmath}
\usepackage{amssymb}
\usepackage{epstopdf}
\usepackage{dcolumn}

\begin{document}

\title{Crossover from inelastic magnetic scattering of Cooper pairs to spin-wave dispersion produces low-energy kink in cuprates}

\author{Tanmoy Das$^{1,2}$, R. S. Markiewicz$^2$ and A. Bansil$^2$}
\address{$^1$ Theoretical Division, Los Alamos National Laboratory, Los Alamos, New Mexico 87544, USA\\ $^2$ Physics Department, Northeastern University, Boston, Massachusetts, 02115, USA}
\date{\today}
\begin{abstract}
We present GW based self-energy calculations for the state of coexisting spin-density wave and $d$-wave superconductivity in a series of cuprate superconductors. In these systems, the spin resonance spectrum exhibits the typical `hour-glass' form, whose upward and downward dispersion branches come from the gapped spin-wave and magnetic scattering of Cooper pairs, respectively. We show that the crossover between these two different dispersion features leads to an abrupt change in slope in the quasiparticle self-energy, and hence the low-energy kink commences in the single-particle quasiparticle spectrum. The calculated electron-bosonic coupling strength agrees well with experimental data as a function of temperature, doping and material. The results demonstrate that the electronic correlations dominate the quasiparticle spectra of cuprates near the low-energy kink, suggesting a relatively smaller role for phonons in this energy range.
\end{abstract}
\maketitle \narrowtext

\section{introduction}

Fundamental information for quantifying the strength of correlation effects and their role in magnetic and superconducting (SC) properties comes from the determination of the magnitude of quasiparticle renormalization, and its intriguing relationship with the quasiparticle line-shape. In conventional superconductors, the most convincing demonstration that electron-phonon coupling is responsible for both dispersion anomalies and superconductivity was obtained from the quantitative correspondence between features in the electronic tunnelling conductance and the phonon spectrum measured by inelastic neutron scattering.\cite{parks} If unconventional superconductivity is mediated by exotic bosons  a similar correspondence should hold, even though details of the coupling between these excitations and the quasiparticle state may vary. In cuprates, pnictide and heavy-fermion superconductors, the emergence of the SC state near the end-point of static antiferromagnetic (AFM) N\'eel order makes these materials among the most promising contenders to realize spin-fluctuation mediated pairing.\cite{HF_sato,actinide,TDahm,inosov} In heavy fermion systems, coupling of the spin fluctuations to the quasi-localized rare-earth $f$-electrons washes out signatures of the dispersion anomaly and complicates the analysis. Similarly, in iron-based superconductors, the presence of multiple bands, multiple gaps and possibly multiple `hot-spots'\cite{Dastworesonance} makes such comparative study difficult and ambiguous.

On the other hand, single band cuprates should provide a clean system for understanding the microscopic origin of characteristic quasiparticle anomalies in their two-particle spectra. In particular, the experimental observation of sudden changes or `kinks' in the quasiparticle dispersion near $50-70$meV in cuprate superconductors\cite{lanzara,kordyuk,WZhang,schmitt,lanzarak,lekx,lekx2,lekx_lsco} has raised the hope that the bosonic excitations responsible for these kinks might also mediate electron pairing in these materials. However, the kink appears at an energy scale where phonons,\cite{lanzara,lee} polarons,\cite{polaron} plasmons,\cite{bob_plasmon} and spin-fluctuations\cite{TDahm} can all contribute to its origin. In fact, significant isotope effects,\cite{lanzara} lattice coupling,\cite{polaron} and charge and spin fluctuations\cite{braicovich,vignolle,tranquada} have all been reported in the kink energy range. Furthermore, poorly understood phenomena associated with underdoping, such as the normal state pseudogap,  competition of superconductivity with incommensurate stripe physics, and nematic order which opens a gap over this energy scale make this problem more complex. The fundamental problem underlying this complexity comes from a lack of consistency between various spectroscopies which are sensitive to different aspects of the ground state, which has hitherto precluded the development of a commonly accepted theory.

Here we go beyond these earlier postulates by performing realistic calculations of the electronic excitation spectra including all channels of spin and charge degrees of freedom in the coexisting spin-density wave (SDW) and superconducting (SC) ground state.\cite{tanmoyop} The calculations reproduce the typical hourglass shape of the magnetic susceptibility, as revealed by inelastic neutron scattering.\cite{Dasresonance} We find a new interpretation for this form, with the lower branch associated with scattering of the Bogolyubov quasiparticles and the upper branch corresponding to a gapped spin wave spectrum. These two branches  meet to form a magnetic resonance peak at a material specific energy.  In turn this peak interacts with the electronic dispersion via the GW self energy, causing an abrupt break in the dispersion known as the `low-energy kink' (LEK).  These low-energy bosons are qualitatively distinct from the strongly-correlated paramagnons responsible for the `high-energy kink'.\cite{MSB} We present the doping, temperature, and momentum dependences of the LEK for Bi$_2$Sr$_2$CaCu$_2$O$_{8}$ (Bi2212), La$_{2-x}$Sr$_x$CuO$_4$ (LSCO) and
Nd$_{2-x}$Ce$_x$CuO$_4$ (NCCO). Significantly, the present results are consistent with the bosonic spectra found earlier in neutron scattering, resonant inelastic x-ray spectroscopy (RIXS), Raman scattering, and studies of the optical `glue' function, as well as with the excitations responsible for the high-energy kink seen in angle-resolved photoemission spectroscopy (ARPES).

This paper is organized as follows. In Sec.~II, we describe the calculation of the GW self-energy due to spin and charge fluctuations in the state of coexisting SDW and $d$-wave superconductivity. In Sec.~IIIA, we describe the microscopic origin of the LEK. The calculated single-particle spectra in the LEK region are compared with experimental data in Sec.~IIIB. The momentum, temperature and doping dependence of the LEK are given in Sec.~IIIC. The corresponding values of the electron-boson coupling constant are compared with experiments in Sec.~IIID. Finally, we conclude in Sec.~IV. Some technical details are found in the Appendices.

\section{Formalism}

 Our starting Hamiltonian with competing Hubbard interaction and
$d-$SC order is\cite{tanmoyop}
\begin{eqnarray}\label{Ham}
H&=&\sum_{{\bf k},\sigma}(\epsilon_{{\bf k}}-\epsilon_F)c^{\dag}_{{\bf k},\sigma}
c_{{\bf k},\sigma}+U\sum_{{\bf k},{\bf k}^{\prime}}c^{\dag}_{{\bf k}+{\bf Q},\uparrow}
c_{{\bf k},\uparrow}c^{\dag}_{{\bf k}^{\prime}-{\bf Q},\downarrow}
c_{{\bf k}^{\prime},\downarrow}\nonumber\\
&&~~+\sum_{{\bf k},{\bf k}^{\prime}}V({\bf k},{{\bf k}^{\prime}})c^{\dag}_{{\bf k},\uparrow}
c^{\dag}_{-{\bf k},\downarrow}c_{-{\bf k}^{\prime},\downarrow}c_{{\bf k}^{\prime},\uparrow}
\end{eqnarray}
where $c^{\dag}_{{\bf k},\sigma} (c_{{\bf k},\sigma})$ is the
electronic creation (destruction) operator with momentum ${\bf k}$
and spin $\sigma=\pm$, $\epsilon_{\bf k}$ is the free particle
dispersion, taken from a tight-binding parametrization of the first-principles band structure with no adjustable parameters, [obtained values are listed in Table ~\ref{Tab1}], and $\epsilon_F$ is the chemical potential.
The quadratic terms are expanded within Hartree-Fock formalism, and the
$d-$wave SC gap is calculated using BCS formalism as
$\Delta_{k}=Vg_k\sum_{k^{\prime}}g_{k^{\prime}} \big<
c^{\dag}_{{\bf k^{\prime}},\uparrow}c^{\prime}_{{-\bf
k^{\prime}},\downarrow}\big>$, where the $d$-wave structure factor is $g_k=\cos{k_xa}-\cos{k_ya}$.
The average is taken over the ground state with combined SC and spin-density-wave (SDW) order.
Here the pairing interaction $V$ is taken to be a momentum independent
parameter which gives the experimental value of the
SC gap $\Delta$ at $T=0$ and $V({\bf k},{{\bf k}^{\prime}})=Vg_kg_{k'}$.
Similarly, the pseudogap is taken as $US$, where $S$ is
the self-consistent mean-field SDW order parameter
$S=\sum_{\bf k}\big<\sigma c^{\dag}_{{\bf k+Q},\sigma}c_{{\bf
k},\sigma}\big>$ at the SDW nesting vector $Q=(\pi,\pi)$. With this, the Hamiltonian in Eq.~\ref{Ham} can be diagonalized straightforwardly and the resulting quasiparticle dispersion consists of upper ($\nu=+$) and lower ($\nu=-$) magnetic bands (U/LMB) further split by superconductivity:
\begin{eqnarray}\label{eigen}
E_{\bf k}^{\nu}=\pm\big(\left(E^{s,\nu}_{k}\right)^2+\Delta_{k}^2\big)^{1/2}.
\end{eqnarray}
Here $E_{\bf k}^{s,\nu}=\xi_{\bf k}^++\nu E_{0k}$ is the quasiparticle dispersion in the non-superconducting SDW state, $E_{0{\bf k}}=\big[\left(\xi_{\bf k}^-\right)^2+(US)^2\big]^{1/2}$ and $\xi_{\bf k}^{\pm}=[\xi_{\bf k}\pm\xi_{{\bf k}+{\bf Q}}]/2$.

The unit cell doubling in the SDW state causes the correlation functions, such as the Lindhard susceptibility, to become tensors, with off diagonal terms in momentum space representation associated with umklapp processes at ${\bf Q}$\cite{schrieffer,chubukov}. In the SDW state, charge- and longitudinal susceptibilities become coupled at finite doping\cite{schrieffer}. In common practice the transverse, longitudinal spin and charge susceptibilities are denoted as $\chi^{+-}, \chi^{zz}$ and $\chi^{\rho\rho}$ respectively. We collect all the terms into a single notation as $\chi^{\sigma\bar{\sigma}}$ where $\bar{\sigma}=\sigma$ gives the
charge and longitudinal components and $\bar{\sigma}=-\sigma$ stands for the transverse component. The noninteracting Lindhard susceptibility in the SDW-BCS case is a $4\times4$ matrix whose components are\cite{schrieffer}
%
%\begin{widetext}
\begin{eqnarray}
\chi_{ij}^{\sigma\bar{\sigma}}({\bf q},\omega)
&=&\frac{1}{N\beta}\sum_{{\bf k},n,s}%\sum_n\sum_s
G_{is}({\bf k},\sigma,i\omega_n)G_{sj}({\bf k}+{\bf q},\bar{\sigma},i\omega_n+\omega)\label{chi1}\nonumber\\
&&\\
&=&\frac{1}{N}\sum_{{\bf k},\nu\nu^{\prime}}^{\prime}A^{\sigma\bar{\sigma}}_{\nu\nu^{\prime},ij}
\sum_{m=1}^3 C^{m}_{\nu\nu^{\prime}}\chi^{m}_{\nu\nu^{\prime}}({\bf k},{\bf q},\omega).
\label{chi2}
\end{eqnarray}
%\end{widetext}
%
We obtain Eq.~\ref{chi2} from Eq.~\ref{chi1} after performing the Matsubara summation over $n$.
$G$ is the $4\times 4$ single-particle Green's function constructed from Eq.~\ref{Ham} in the Nambu space.
The summation indices $\nu (\nu^{\prime}) = \pm$ gives upper and lowe magnetic bands, respectively.
Here, the coherence factor due to SDW order in the particle-hole channel is
\begin{eqnarray}\label{chiSDW}
A^{\sigma\bar{\sigma}}_{\nu\nu^{\prime},11/22}&=&\frac{1}{2}\left(1\pm\nu\nu^{\prime}\frac{\xi_{\bf k}^-\xi_{{\bf k}+{\bf q}}^-+\sigma\bar{\sigma}(US)^2}{E_{0{\bf k}}E_{0{\bf k}+{\bf q}}}\right),\nonumber\\
%
%A^{SDW,\sigma\bar{\sigma}}_{\nu\ne\nu^{\prime},11/22}&=&\frac{1}{2}\left(1\mp\frac{\xi_{\bf k}^-\xi_{{\bf %k}+{\bf q}}^-+\sigma\bar{\sigma}(US)^2}{E_{0{\bf k}}E_{0{\bf k}+{\bf q}}}\right),\nonumber\\
%
A^{\sigma\bar{\sigma}}_{\nu\nu^{\prime},12/21}&=&-\nu\frac{US}{2}\left(\frac{\sigma}{E_{0{\bf k}}}+\nu\nu^{\prime}\frac{\bar{\sigma}}{E_{0{\bf k}+{\bf q}}}\right).
%
%A^{SDW,\sigma\bar{\sigma}}_{\nu\ne\nu^{\prime},12/21}&=&-\nu\frac{US}{2}\left(\frac{\sigma}{E_{0{\bf %k}}}-\frac{\bar{\sigma}}{E_{0{\bf k}+{\bf q}}}\right).
%
\end{eqnarray}
The SC coherence factors are
\begin{eqnarray}\label{t11}
C^{\rm 1}_{\nu\nu^{\prime}}
%%%%%
&=&\frac{1}{2}\left(1+\frac{E^{s,\nu}_{\bf{k}}
E^{s,\nu^{\prime}}_{\bf{k}+\bf{q}}+\Delta_{\bf{k}}\Delta_{\bf{k}+\bf{q}}}
{E^{\nu}_{\bf{k}}E^{\nu^{\prime}}_{\bf{k}+\bf{q}}}\right),\nonumber\\
%%%%%%
C^{\rm 2/3}_{\nu\nu^{\prime}}&=&\frac{1}{4}\left(1\pm\frac{E^{s,\nu}_{\bf{k}}}{E^{\nu}_{\bf{k}}}
\mp\frac{E^{s,\nu^{\prime}}_{\bf{k}+\bf{q}}}{E^{\nu^{\prime}}_{\bf{k}+\bf{q}}}%\right.\nonumber\\
%
%&&~~~~~~\left.
-\frac{E^{s,\nu}_{\bf{k}}E^{s,\nu^{\prime}}_{\bf{k}+\bf{q}}+\Delta_{\bf{k}}\Delta_{\bf{k}+\bf{q}}}
{E^{\nu}_{\bf{k}}E^{\nu^{\prime}}_{\bf{k}+\bf{q}}}\right).\nonumber\\
\label{sccoherence}
\end{eqnarray}
Lastly the index $m$ represents the summation over three polarization bubbles
related to the quasiparticle scattering ($m=1)$, quasiparticle pair
creation ($m=2$) and pair annihilation ($m=3$), as defined by
\begin{eqnarray}
\chi^{1}_{\nu,\nu^{\prime}}({\bf k},{\bf q},\omega)&=&
-\frac{f(E^{\nu}_{{\bf k}})-f(E^{\nu^{\prime}}_{{\bf k}+{\bf q}})}
{\omega+i\delta+(E^{\nu}_{{\bf k}}-E^{\nu^{\prime}}_{{\bf k}+{\bf q}})},
\label{Eq:ch3_chi_0_sc1}\\
\chi^{2,3}_{\nu,\nu^{\prime}}({\bf k},{\bf q},\omega)&=&\mp
\frac{1-f(E^{\nu}_{{\bf k}})-f(E^{\nu^{\prime}}_{{\bf k}+{\bf q}})}
{\omega+i\delta\mp(E^{\nu}_{{\bf k}}+E^{\nu^{\prime}}_{{\bf k}+{\bf q}})}.
\label{Eq:ch3_chi_0_sc23}
\end{eqnarray}
It is interesting to observe that $\chi^{1}$ is the particle-hole scattering term which becomes gapped in the SC state while the pair scattering terms, $\chi^{2,3}$, only contribute in this region. We show below that this crossover from the normal to the superconducting state gives rise to the low-energy kink.

In the normal state, the $2\times2$ RPA susceptibility is obtained from the standard formula\cite{schrieffer}
\begin{widetext}
\begin{eqnarray}\label{RPASpinSucscomp}
\chi_{RPA,11}^{\sigma\bar{\sigma}}({\bf q},\omega)
&=&\frac{\bigl[1+\sigma\bar{\sigma} U\chi_{22}^{\sigma\bar{\sigma}}({\bf q},\omega)\bigr]\chi_{11}^{\sigma\bar{\sigma}}({\bf q},\omega) +
U\bigl[\chi_{12}^{\sigma\bar{\sigma}}({\bf q},\omega)\bigr]^2}
{\bigl[1-U\chi_{11}^{\sigma\bar{\sigma}}({\bf q},\omega)\bigr]\bigl[1+\sigma\bar{\sigma}U\chi_{22}^{\sigma\bar{\sigma}}({\bf q},\omega)\bigr]
+\sigma\bar{\sigma}\bigl[U\chi_{12}^{\sigma\bar{\sigma}}({\bf q},\omega)\bigr]^2},
%%%
\label{RPASpinSucscomp11}\\
%%%
\chi_{RPA,22}^{\sigma\bar{\sigma}}({\bf q},\omega)
 &=&\frac{\bigl[1-U\chi_{11}^{\sigma\bar{\sigma}}({\bf q},\omega)\bigr]\chi_{22}^{\sigma\bar{\sigma}}({\bf q},\omega) + U\bigl[\chi_{12}^{\sigma\bar{\sigma}}({\bf q},\omega)\bigr]^2}
{\bigl[1-U\chi_{11}^{\sigma\bar{\sigma}}({\bf q},\omega)\bigr]\bigl[1+\sigma\bar{\sigma}U\chi_{22}^{\sigma\bar{\sigma}}({\bf q},\omega)]
+\sigma\bar{\sigma}\bigl[U\chi_{12}^{\sigma\bar{\sigma}}({\bf q},\omega)\bigr]^2},
%
%%%
\label{RPASpinSucscomp22}\\
%%%
\chi_{RPA,12/21}^{\sigma\bar{\sigma}}({\bf q},\omega)
 &=&\frac{\chi_{12}^{\sigma\bar{\sigma}}({\bf q},\omega)}
{\bigl[1-U\chi_{11}^{\sigma\bar{\sigma}}({\bf q},\omega)\bigr]\bigl[1+\sigma\bar{\sigma}U\chi_{22}^{\sigma\bar{\sigma}}({\bf q},\omega)]
+\sigma\bar{\sigma}\bigl[U\chi_{12}^{\sigma\bar{\sigma}}({\bf q},\omega)\bigr]^2}.
%
%%%
\label{RPASpinSucscomp12}
%%%
\end{eqnarray}
\end{widetext}
In the longitudinal and charge channel ($\bar{\sigma}=\sigma$, denoted by subscript $`cz$'), the RPA corrections do not introduce any new poles and thus all the normal state structure lies above the charge gap in the particle-hole continuum. Along the transverse direction ($\bar{\sigma}=-\sigma$, denoted by subscript `$t$'), a linear spin-wave dispersion develops in the normal state which extends to zero energy at $Q$.\cite{schrieffer} The necessary condition to yield a gapless Goldstone mode is that Eqs.~\ref{RPASpinSucscomp11}-\ref{RPASpinSucscomp12} reduce to the self-consistent SDW order parameter at $q=Q$, which is indeed the case in the normal state.

In the SC state, this zero energy spin-wave shifts to $\omega=|\Delta_{{\bf k}_F}|+|\Delta_{{\bf k}_F+{\bf Q}}|$, due to the particle-particle (and hole-hole) scattering terms $\chi^{2,3}$ in Eq.~\ref{Eq:ch3_chi_0_sc23}. These terms have finite intensity only if the SC gap changes sign at the `hot-spot' ${\bf Q}$,\cite{chubukov_res,eremin_res} see Eq.~\ref{sccoherence}. Above the SC gap, the spin-wave term coming from Eq.~\ref{Eq:ch3_chi_0_sc1} is turned on. We show below that this crossover produces the characteristic hour-glass phenomenology and is ultimately responsible for the LEK.

In a GW formalism, the $W$ describes the dressing of the Green's function $G$ by electron-hole pairs acting as bosonic modes.  In our purely electronic model we calculate the latter, which couple to the electrons causing an increase in the low-energy electronic mass and a shortening of the lifetime at higher energies. These effects are described in terms of a complex self-energy $\Sigma$ which is calculated within the $GW$ approximation to Feynman-Dyson perturbation  theory\cite{tanmoyop} as
\begin{eqnarray}
&\Sigma_i({\bf k},\sigma,i\omega_n)=\frac{1}{2}\eta_{i}{\bar U}^2
\sum_{{\bf q},\sigma^{\prime}}
\int_{0}^{\infty}\frac{d\omega_p}{2\pi}\nonumber\\
&G({\bf k}+{\bf q},\sigma^{\prime},i\omega_n,\omega_p)
\Gamma({\bf k},{\bf q},i\omega_n,\omega_p)\chi^{\prime\prime}_{i}({\bf
q},\sigma^{\prime},\omega_p),%\nonumber\\
\label{selfenergy}
\end{eqnarray}
Here $\Gamma$ is the vertex correction, modeled using Ward's identity. Since the bosons are `built' from fermions, self-consistency is required in calculating the self-energy, as in the quasiparticle-$GW$ approximation developed in Refs.~\protect\onlinecite{tanmoyop,QPGW,NFL}.

\section{Results}

\subsection{Microscopic origin of low-energy kink}

\begin{figure}[top]
%\hskip-1in
%\rotatebox{0}{\scalebox{0.7}{\includegraphics{chi_lsco.eps}}}
\rotatebox{0}{\scalebox{0.5}{\includegraphics{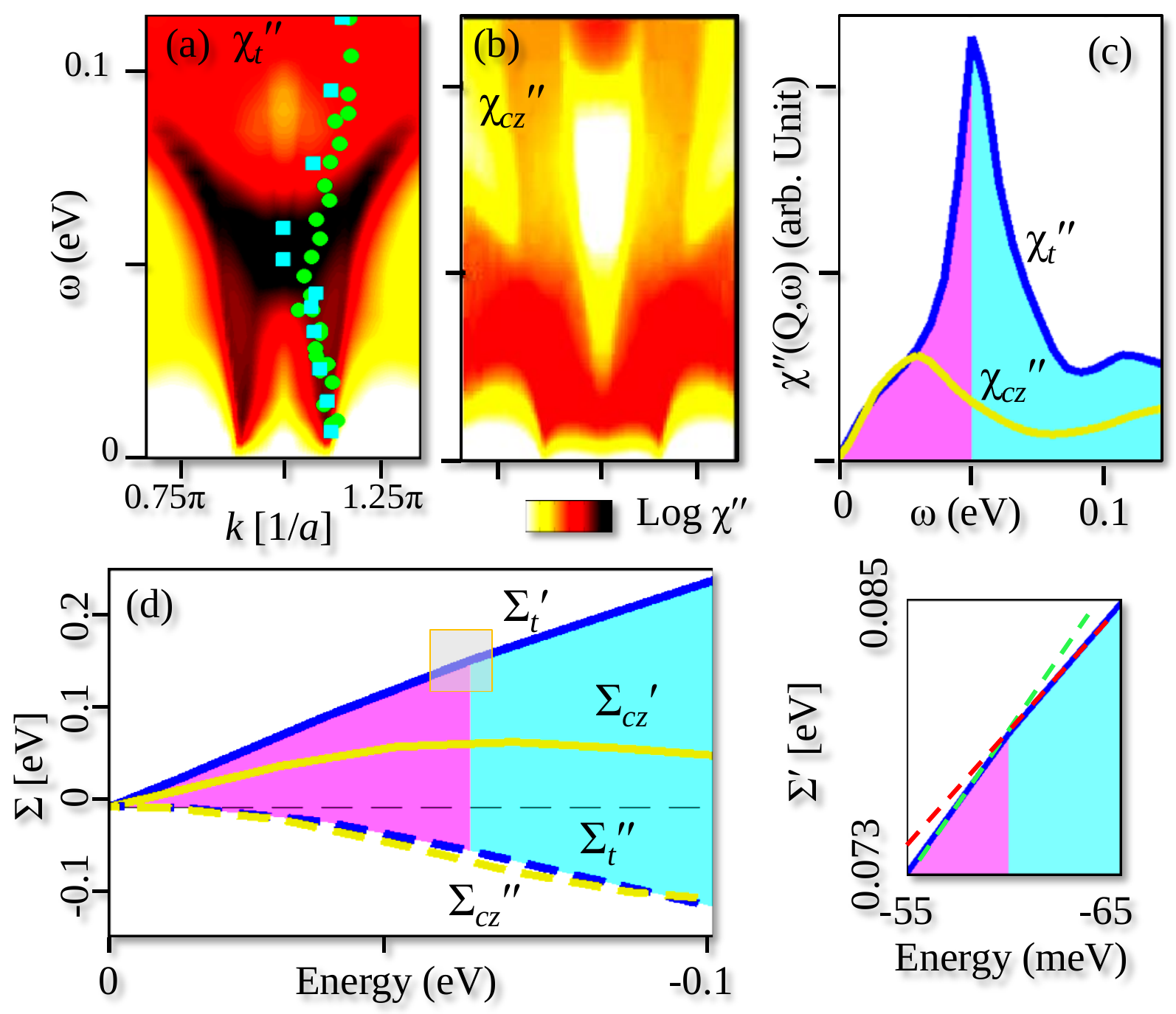}}}
\caption{(Color online) (a) Transverse susceptibility $\chi_t^{\prime\prime}$ is plotted along the diagonal direction centered at ${\bf Q}=(\pi,\pi)$ for LSCO at $x=0.16$.  Symbols are corresponding neutron data for LSCO at $x=0.16$ (circles)\cite{vignolle} and LBCO at $x=1/8$ (squares)\cite{tranquada}. (b) Same as (a) but in the longitudinal + charge channel for LSCO at $x=0.16$. (c) $\chi^{\prime\prime}$ at ${\bm Q}$ as a function of energy. (d) Corresponding quasiparticle self-energy for both transverve ($\Sigma_t$) and longitudinal + charge channel ($\Sigma_{cz}$). The magenta and cyan shadings differentiate the SC scattering and spin-wave spectra. The adjacent {\it inset} expands the energy scale to clarify the nature of crossover energy scales present in the corresponding left hand side figure.
}\label{fig1}
\label{alp2F}
\end{figure}

In Figures \ref{fig1}(a) and (b), we present the imaginary part of the transverse and longitudinal + charge channels of the RPA susceptibilities (denoted by $\chi_t^{\prime\prime}$, and $\chi_{cz}^{\prime\prime}$, respectively) near ${\bm Q}=(\pi,\pi)$ along the diagonal direction, for LSCO at one representative doping $x=0.16$. The momentum integrated values are given in Fig.~\ref{fig1}(c).
%The compared with a variety of experimental determinations\cite{guarise,braicovich,vignolle,tranquada}.
More extensive data for the whole Brillouin zone and over a large energy range and also for Bi2212 and NCCO are presented in Fig.~\ref{fig6} below. As we have shown in our earlier calculations, the normal state spin-wave dispersion becomes massive near ${\bm Q}$ due to SC gap opening.\cite{Dasresonance} Below this spin gap, the Bogolyubov scattering of the SC states on the Fermi surface gives an oppositely dispersive branch, leading to an hour-glass dispersive feature. %$\chi_t^{\prime\prime}$ at ${\bm Q}$ is plotted in Fig.~\ref{fig1}(b) which shows a sharp resonance peak. The origin of the peak that at the energy the upward and down-wave dispersion of different origin meet.\cite{Dasresonance}

The real part of the computed self-energy $\Sigma^{\prime}$ is plotted of Fig.~\ref{fig1}(c) in the low-energy region. Above this energy scale the self-energy becomes dominated by normal state paramagnon contributions which are important for the study of the high-energy kink,\cite{MSB} but not in the present case. The origin of the low-energy kink is different. At low energy, $\Sigma^{\prime}$ shows a break in slope [see {\it inset} figure], rather than a peak, which corresponds to the neutron mode discussed above.\cite{foot_mode_shift} This break in slope in $\Sigma^{\prime}$, occurring near $50-70$meV depending on the material under study, leads to the LEK.

\subsection{Single particle dispersion and the LEK}

\begin{figure}[top]
%\hskip-2in
%\rotatebox{270}{\scalebox{0.55}{\includegraphics{figure_lek_lsco_bi2212_ncco.eps}}}
\rotatebox{0}{\scalebox{0.6}{\includegraphics{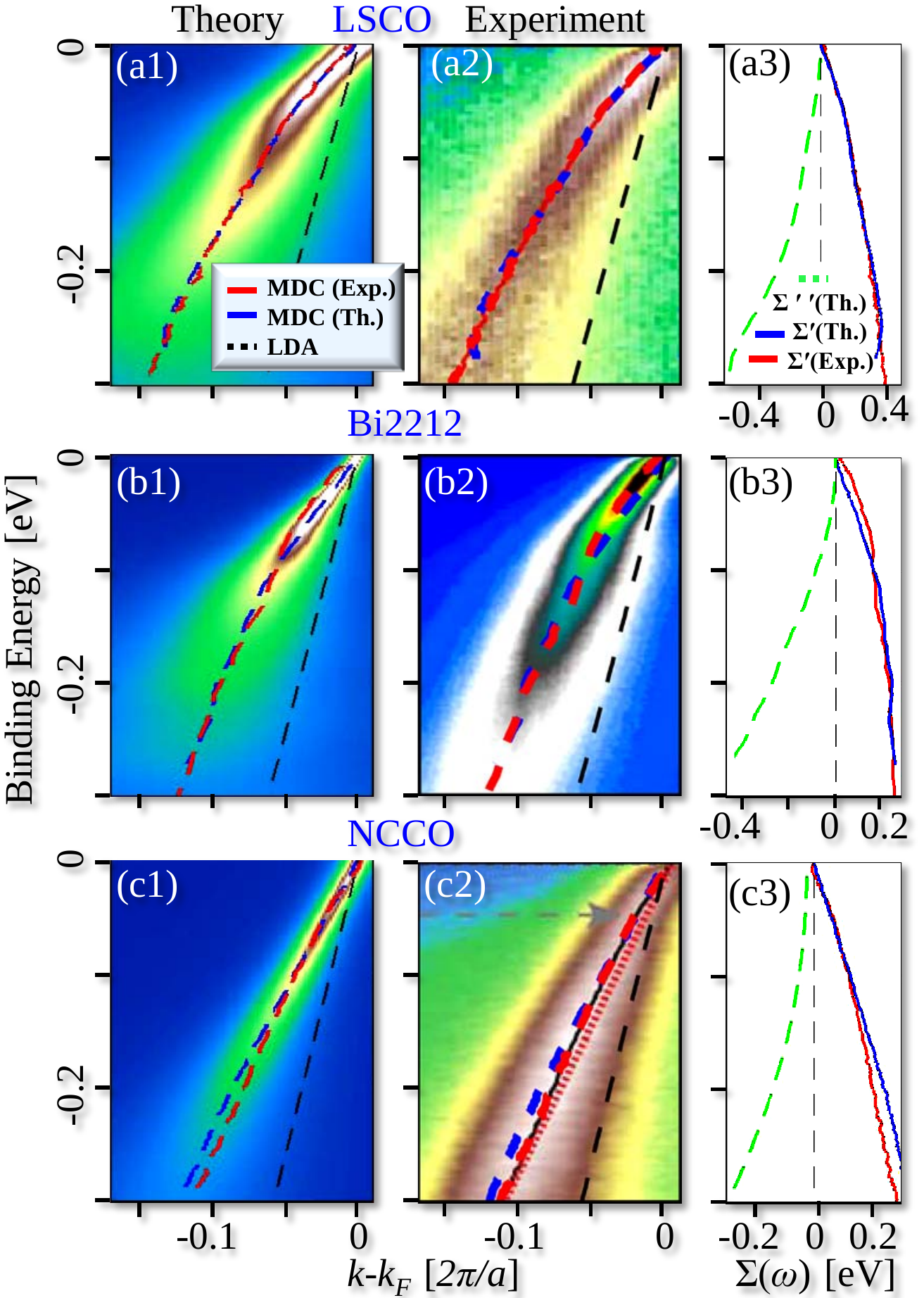}}}
\caption{%{\bf Low energy kink for LSCO, Bi2212 and NCCO at optimal doping.}
(Color online) (a1) Theoretical and (a2) experimental\cite{kordyuk} single-particle dispersion in the energy range of the LEK for LSCO. The blue and green dashed lines give the MDC peak positions for the
theoretical and experimental spectra, while the non-interacting LDA dispersion is plotted as a red
dashed line. (a3) The real part of the self-energy, calculated as the difference between
the MDC peaks and the LDA dispersion. Dashed green line gives our computed imaginary part of the self-energy.
Similar results are shown for Bi2212 [Ref.~\protect\onlinecite{WZhang}] in (b1,b2,b3) and for NCCO [Ref.~\protect\onlinecite{schmitt}] in (c1,c2,c3).} \label{fig2}
\end{figure}

The self-energy dressed single-electron Green's function is $G_d^{-1}=G^{-1}-\Sigma$, where the Green' function and self-energies are 4$\times$4 matrices as defined above.\cite{tanmoyop} The single particle spectrum is then computed as $A({\bm k},\omega)=- {\rm Im}G_{d,11}({\bm k},\omega)/\pi$, and plotted along the nodal directions in Fig.~\ref{fig2} for the three materials LSCO, Bi2212, and NCCO.\cite{foot_mode_shift} The corresponding experimental data\cite{kordyuk,WZhang,schmitt} are given in the adjacent middle column. Theory and experiment demonstrate good agreement both in the shape of the dispersion and in the associated spectral weight. Notably, our calculation neglects possible modulations of the spectral weight due to matrix element effects, which can be important in ARPES\cite{arpesab}, STM\cite{joukostm}, inelastic light scattering\cite{MBRIXS}, and other spectroscopies\cite{comptonab,positronab}.
For ease in comparison, the dashed lines in both figures show the dispersion calculated as the position of the peaks in the theoretical (blue) and experimental (red) momentum distribution curves (MDCs). [The MDC is defined as a plot of $A(k,\omega)$ vs momentum $k$ at fixed energy $\omega$.] The real part of the self-energy $\Sigma^{\prime}$ is typically determined as the difference between the MDC peaks and the bare LDA dispersion (black dashed lines). We plot the corresponding experimental and theoretical estimates of $\Sigma^{\prime}$ in the right column of Fig.~\ref{fig2}, which also displays the theoretical values of $\Sigma^{\prime\prime}$. For single layer systems, the LEK is around 70~meV for LSCO, but 50~meV for NCCO both in theory and experiment while for Bi2212 our theory finds a larger value of the kink energy around 100~meV whereas experimental data show a kink near 70~meV.

\subsection{Momentum, temperature and doping dependence of LEK}

\begin{figure}[top]
%\hskip-2in
\rotatebox{0}{\scalebox{0.6}{\includegraphics{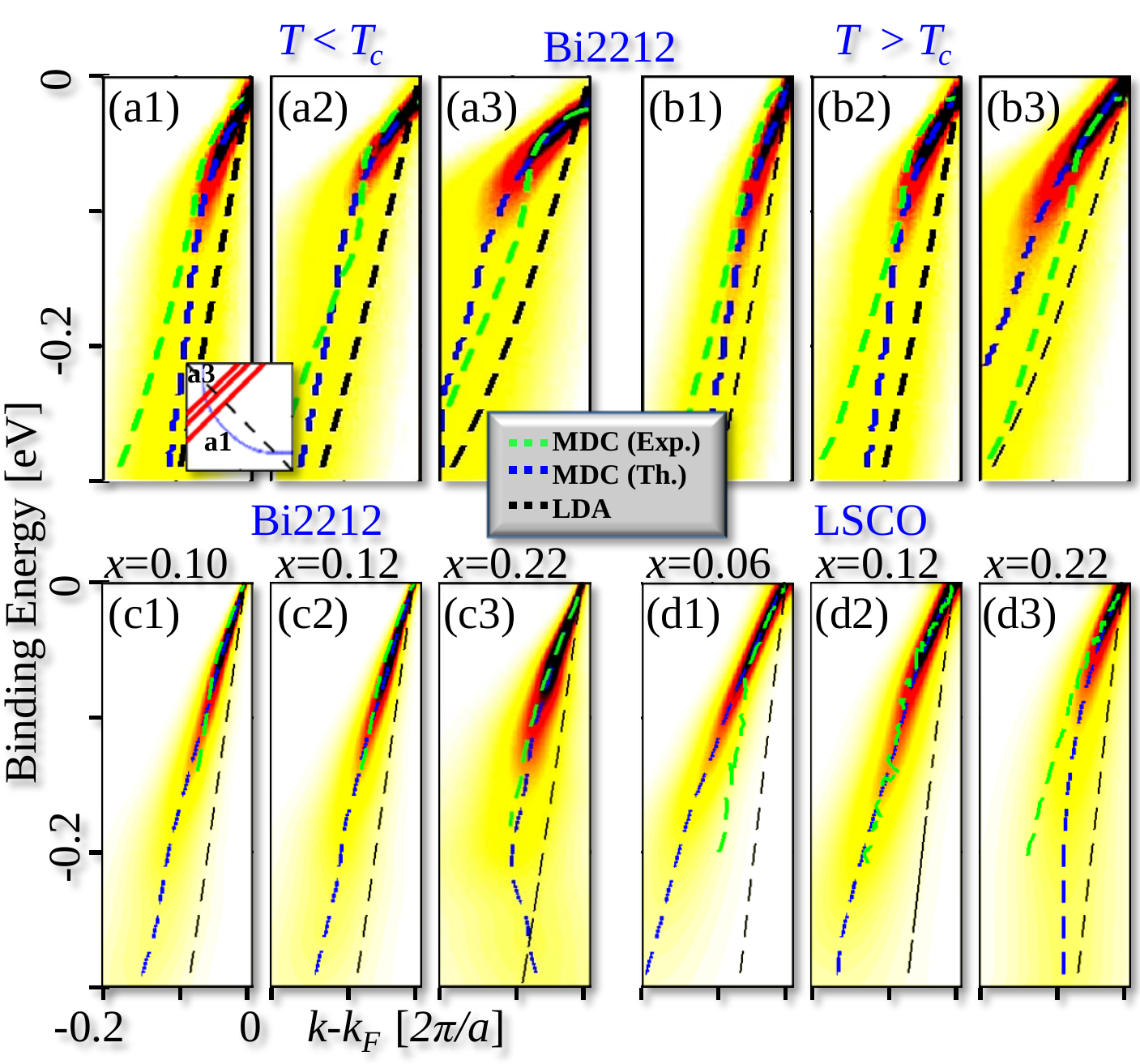}}}
\caption{(Color online) (a1-a3) and (b1-b3): Momentum dependence of the LEK in optimally doped Bi2212 at both $T<T_c$ and $T>T_c$. The ARPES data (green line) are taken from Ref.~\onlinecite{lanzarak} at the same momentum. (c1-c3) Doping dependence of the LEK in Bi2212. Corresponding experimental data\cite{lanzara,lekx,lekx2} are plotted by comparing $T_c$ with universal dome feature as given in Appendix~A. (d1-d3) Same as in (c1-c3) but for LSCO where the corresponding experimental data are obtained from Ref.~\onlinecite{lekx_lsco}.}
\label{fig3}
\end{figure}

In Fig.~\ref{fig3} we study the momentum ($k$), temperature ($T$), doping and material dependence of the LEK. At each $T$ and doping, both the SDW and SC gaps are evaluated self-consistently assuming $T$ independent values of $U$ and pairing interaction $V$. The calculated single-particle spectra yield a very good description of the ARPES data over the entire Brillouin zone.
Some discrepancies are noticeable in the antinodal region where the pseudogap and SC gaps are the largest. Here, our
theory slightly overestimates the experimental kink, which is also the case when the pseudogap increases with underdoping (see lower panel). Also, in Bi2212 we have neglected the bilayer splitting of the CuO$_2$ bands, which is largest in
the antinodal direction.  Note that the agreement with experiment would not be significantly improved by including a phonon contribution to the kink.\cite{giustino}

\subsection{Coupling constant}

\begin{figure}[top]
%\hskip-2in
\rotatebox{0}{\scalebox{0.4}{\includegraphics{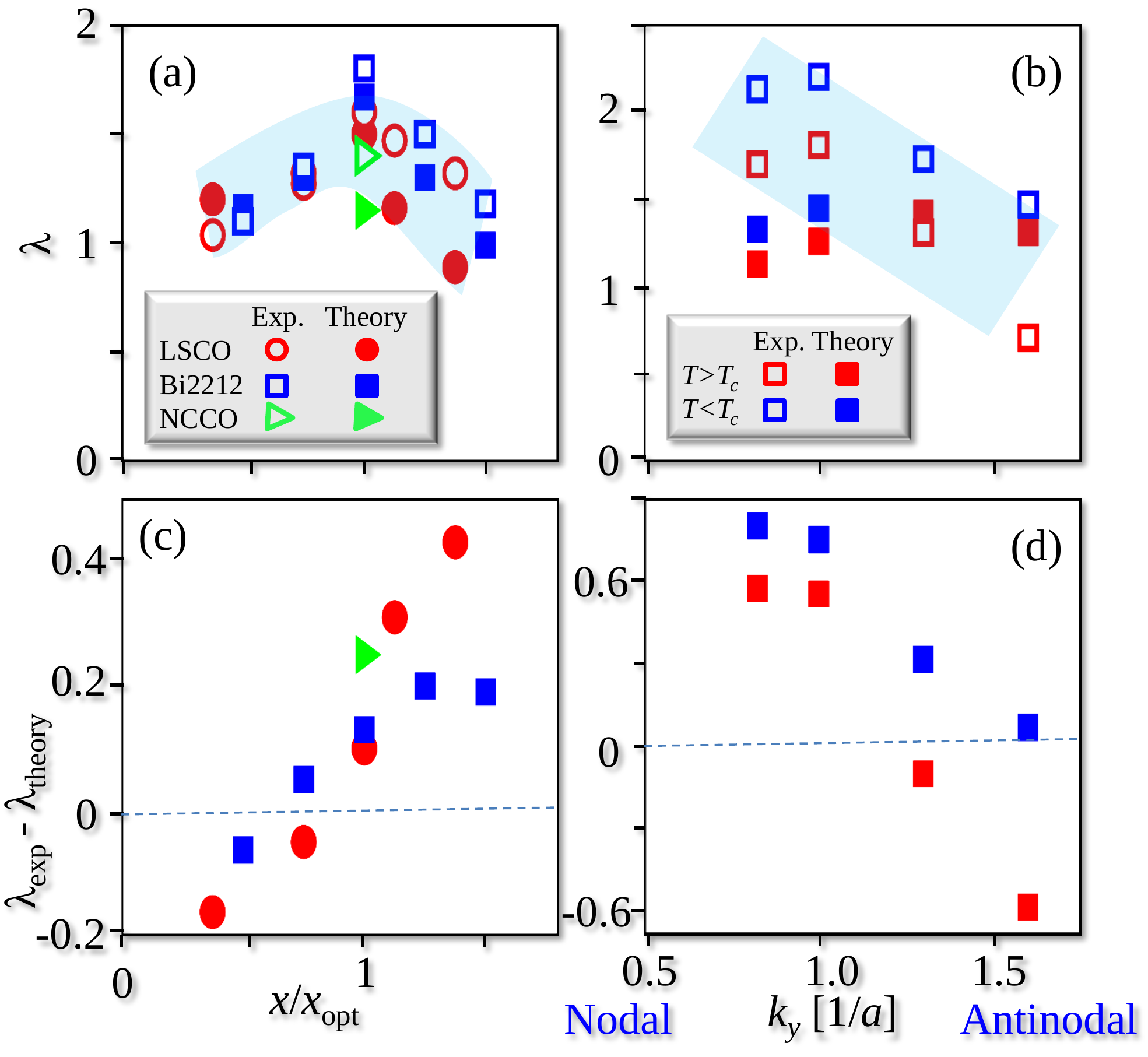}}}
\caption{(Color online) (a) Comparison of experimental and theoretical electron-boson coupling constant $\lambda$ as a function of doping and material. The horizontal scale is normalized to the optimal doping for each material. (b) Same as (a) but as a function of $k$ and $T$ for Bi2212. It is important to distinguish our obtained values of $\lambda$ with most of earlier data [see for example Ref.~\onlinecite{lanzara}]. We calculate $\lambda$ with respect to the actual LDA dispersion whose slope is doping and momentum dependent. In common practice, $\lambda$ is obtained assuming a bare linear dispersion which is often taken to be doping and momentum independent. (c)-(d) Difference of the estimation of $\lambda$ between experiment and theory, using the same symbols as in the corresponding top panels.}
\label{fig4}
\end{figure}

Fig.~\ref{fig4} shows the corresponding electron-boson coupling constant $\lambda$, calculated from the gradient of $\Sigma^{\prime}$ which in the low-energy region becomes $\Sigma'(\xi_{\bf k})=-\lambda \xi_{\bf k}$. $\lambda$ in principle varies with $\omega$ but weakly in the low-energy region below 70~meV and thus we report the values of $\lambda$ at this energy. As shown in Fig.~\ref{fig4}, the experiments and theory agree well in the optimal doping region for all three materials (see figure caption for details). The difference between experimental and theoretical $\lambda$s becomes more prominent as one moves away from the optimal region.  When this difference is positive, it is suggestive of the presence of an additional bosonic coupling, including phonons.  Note that this additional component is always small compared to the main electronic component that we have calculated.

\section{Conclusion}

In summary, we have shown that the low-energy kink arises mainly from a {\it crossover} from Cooper pair scattering to spin-wave dispersion which is different from the paramagnon {\it modes} responsible for high-energy kink (not a crossover).\cite{MSB} The associated electron-boson coupling constant $\lambda$ depends strongly on the slope of the lower dispersion branch of the hourglass, and thus on the FS which leads to a doping, temperature and material dependence of the coupling constant. The susceptibility energy scales are observed directly in the inelastic neutron scattering and optical glue measurements, and should also be seen in RIXS and Raman spectra with improved resolution. We argue that such a crossover feature is a general phenomenon of any coexisting SDW and unconventional SC state and thus should also be present in iron-based and heavy fermion superconductors.

\begin{acknowledgments}
This work is supported by the U.S.D.O.E grant DE-FG02-07ER46352, and benefited from the allocation of supercomputer time at NERSC and Northeastern University's Advanced Scientific Computation Center (ASCC). The work at Los Alamos National Laboratory is funded by US DOE, BES and LDRD.
\end{acknowledgments}

\appendix
\section{Tight-binding parameters and the interaction strength.}

\begin{figure}[top]
\rotatebox{0}{\scalebox{0.3}{\includegraphics{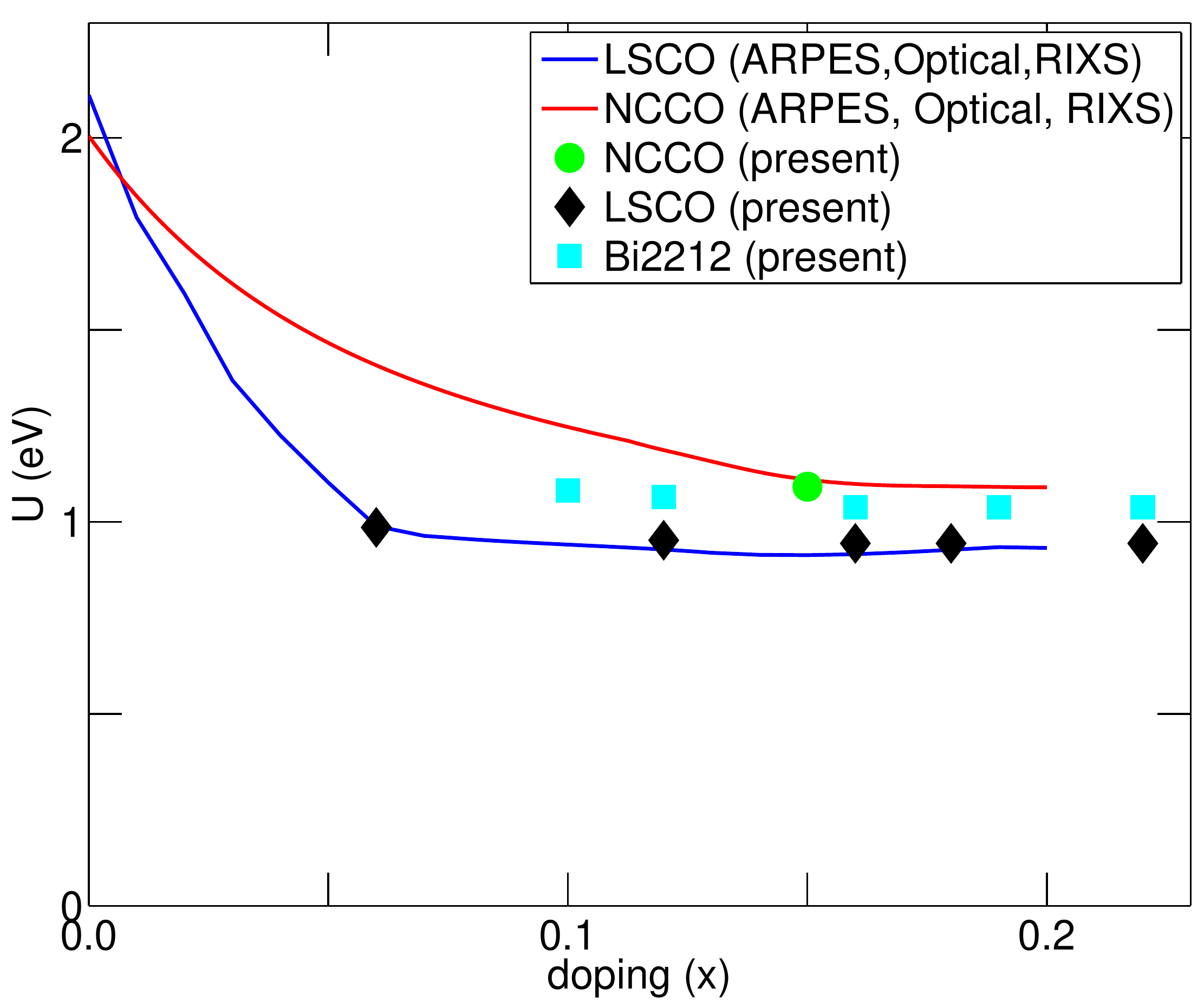}}}
\caption{(Color online) Doping dependent values of $U$. The solid line gives the self-consistent values of $U$ as a function of doping found in the QP-GW model\cite{tanmoyop}. Symbols give the values used in the present calculations for various materials. }
\label{fig5}
\end{figure}

We obtain the tight-binding parameters by fitting to the material specific LDA dispersion\cite{markietb,joukostm,abfoot1,new1,new2} as listed in Table~\ref{Tab1}. The self-energy renormalizes the LDA band to obtain the experimental dispersion. The interaction parameters, both Hubbard $U$s and superconducting pairing potential $V$s, are determined from fits to other experiments -- specifically the values of the superconducting and pseudogaps seen in ARPES, optical and STM etc measurements -- and the resulting values are listed in Table~\ref{Tab2}.\cite{armitage,tanmoyprl,bobtc,yoshida,heufner} The values of $U$ match well with the calculated values of the screened $U,$ as shown in Fig.~\ref{fig5}. Notably, we find that over the doping range $x>0.05$ the bare $U$ is universal -- the same for all materials and dopings. Finally, we find that our computational method also provides a good model for ARPES and optical properties of the cuprates.\cite{tanmoyop,QPGW}

\begin{table}[h]
\centering
\begin{tabular}{|c|c|c|c|c|c|c|c|}
\hline \hline
Material& $t$ & $t^{\prime}$ & $t^{\prime\prime}$& $t^{\prime\prime\prime}$&$Z$\\
\hline
NCCO [Ref.~\onlinecite{markietb}]& 0.42 & -0.1 & 0.065 & 0.0075& 0.4 \\
LSCO [Ref.~\onlinecite{markietb}]& 0.4195 & -0.0375 & 0.018 & 0.034 & 0.3\\
Bi2212 [Ref.~\onlinecite{joukostm}]& 0.44 & -0.1 & 0.05 & 0.0 & 0.4\\
\hline \hline
\end{tabular}
\caption{Tight-binding parameters obtained by fitting to the LDA band-structure with self-consistent renormalization factor $Z$ (right).}
\label{Tab1}
\end{table}
%
%

%\hline \hline
%\newline
%\begin{widetext}
\begin{table*}
\begin{tabular}{|c|c|c|c|c|c|c|c|c|c|}
\hline \hline
Material& Doping ($x$) &$\Delta_{pg}$ (meV)& $U/t$ & $\Delta_{sc}$ (meV)& Pairing Potential & $T_c$ K & $Z$ \\
& & (Exp./Theory) & (Theory)& (Exp./Theory) & $V$ (meV) (Theory)& Exp.(Theory)\\
 \hline
 LSCO & 0.06 & 150 & 2.35 & 6
 & -93 & 18 (27) & 0.5\\
 LSCO &0.12& 120 & 2.27 & 11 & -63 & 30 (48) & 0.48\\
 LSCO &0.16&63  & 2.25 & 15 & -51 & 40 (85) & 0.45\\
 LSCO & 0.18 & 43 & 2.25 & 13
 & -35 & 37 (75) &0.43 \\
 LSCO &0.22&0 & 2.25 & 8  & -28 & 26 (48)&0.4
 \\
 NCCO & 0.15& 170 (at hotspot) [Ref.~\onlinecite{armitage}] & 4.1 &
 5.5 [Ref.~\onlinecite{tanmoyprl}] & -83 & 24 (31) [Ref.~\onlinecite{tanmoyprl}] & 0.4\\
 Bi2212 & 0.10 & 113 & 2.46 & 15
 & -77 & 55 (85) & 0.45 \\
 Bi2212 & 0.12 & 95 & 2.42 & 17.5
 & -75 & 65 (95) & 0.43 \\
 Bi2212 & 0.16 & 75 & 2.36 & 20
 & -67 & 91 (115) & 0.4 \\
 Bi2212 & 0.19 & 50 & 2.36 & 17.5
 & -58 & 70 (90) & 0.39\\
 Bi2212 & 0.22 & 25 & 2.36 & 12.5
 & -50 & 55 (75) & 0.38\\
\hline \hline
\end{tabular}
\caption{
%Experimental gap values and the resulting self-consistent values of the order parameters.
The value of $U/t$ is chosen to reproduce the experimental pseudogap ($\Delta_{pg}$) whose values are presented here along the hot-spot direction in the electron doped case and the antinodal direction for hole doped cuprates LSCO and Bi2212. Similarly, the parameter value of pairing potential $V$ is taken to reproduce the superconducting gap ($\Delta_{sc}$), whose maximum value along the antinodal direction is presented here. Our mean-field calculations overestimate the values of $T_c$, presumably due to the neglect of phase fluctuations\cite{bobtc}.Experimental data for  LSCO  are taken from Ref.~\onlinecite{yoshida} and for Bi2212 from Ref.~\onlinecite{heufner}.}
\label{Tab2}
\end{table*}

%\end{widetext}

\section{Material dependence of the susceptibility and the self-energy}

\begin{figure*}[top]
%\hskip-2in
\rotatebox{270}{\scalebox{0.7}{\includegraphics{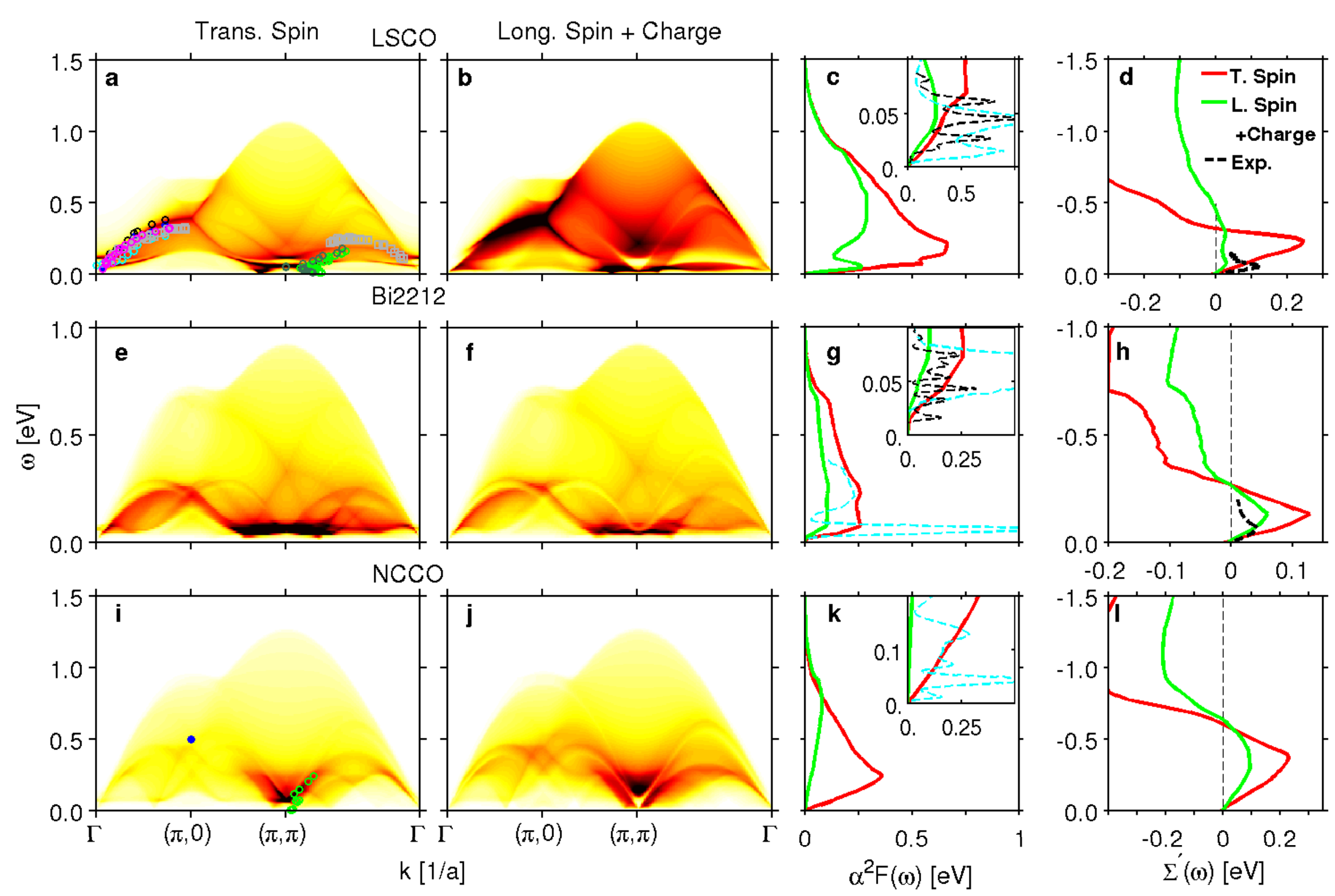}}}
\caption{(Color online) The imaginary part of the susceptibility is plotted along the high-symmetry lines of the Brillouin zone for transverse [first column] and longitudinal plus charge channels [second column] for two hole doped (LSCO and Bi2212) and one electron doped cuprate (NCCO) near their optimal dopings. The results are compared with RIXS data for insulating LCO [blue from Ref~\onlinecite{braicovich}], underdoped LSCO [black and cyan symbols from Ref.~\onlinecite{braicovich}], and  undoped SCOC [grey symbols].\cite{guarise} Magenta and green symbols are neutron data for undoped and optimally doped LSCO\cite{coldea,vignolle} which are compared with the LBCO data at 1/8 doping (deep green symbols)\cite{tranquada}. The corresponding bosonic spectral weight $\alpha^2F(\omega)$ (see text) is compared for transverse (red) and the mixed state of longitudinal spin and charge (green) channels in the third column. The corresponding insets expand the low-energy part of the calculated $\alpha^2 F(\omega)=U^2\int d^2q \chi^{\prime\prime}({\bm q},\omega)$ and compare them with data extracted from optical (cyan dashed lines) and ARPES spectra (black dashed line). The optical data are taken for nearly optimally doped LSCO and for Bi2212 from Refs.~\protect\onlinecite{hwangoplscobi2212,hwang}, and electron doped PCCO from Ref.~\protect\onlinecite{schachinger}. ARPES-derived data are for LSCO at $x=0.03$\cite{zhouZX} and overdoped Bi2212\cite{LZhao}. The corresponding quasiparticle self-energy is presented in the right column. Black dashed lines are ARPES self-energy data from which $\alpha^2F$ was extracted.
}
\label{fig6}
\end{figure*}

Calculated spectra of $\chi^{\prime\prime}$ are presented in the first two columns of the Fig.~\ref{fig6} as a function of excitation energy along high-symmetry directions in momentum space for the transverse (column 1) and longitudinal spin plus charge (column 2) channels; the three rows represent different materials, LSCO, Bi2212 and NCCO. Superimposed on the calculated spectra, we show the single magnon RIXS results for undoped LCO\cite{braicovich}, neutron data of the same sample\cite{braicovich}, and RIXS data of undoped SCOC and NCO.\cite{guarise,coldea,vignolle,tranquada}

We plot the corresponding momentum-averaged Eliashberg, or `glue'  function in the third column of the Fig.~\ref{fig6}.
These functions display several peaks, but the most relevant ones for the present purpose are the two which are present in the energy scales of 300-400meV and $\sim 70$meV for hole doping and $\sim 500$meV and 50meV for electron doping. The high energy peak mainly stems from the strong intensities near $(\pi,0)$ and lies in the `waterfall' or high-energy kink region, while the low-energy peak arises from the magnetic resonance scattering around $(\pi,\pi)$.

The insets in column 3 of Fig.~\ref{fig6} show expanded views of the low-energy region of the $\alpha^2F(\omega)$ functions and compare them with available data from optical absorption\cite{hwangoplscobi2212,hwang,schachinger}
and ARPES (black lines)\cite{zhouZX,LZhao}. In ARPES experiments, the $\alpha^2F$ spectra are usually derived from maximum entropy calculations which tend to produce spectra consisting of a series of peaks,  including the one at $\sim 50-70$meV.  A similar multi-peak structure is seen in the optical-glue function of NCCO (inset)\cite{schachinger}. While both experiments find the feature at $\sim 50-70$meV, there are additional peaks in the low-energy region not reproduced in our calculation which may represent contributions due to phonons or other bosons.  For NCCO, our theory produces a kink rather than a sharp peak at $\sim 50$meV, which is consistent with the weaker strength of the LEK as compared to LSCO and Bi2212, see main Fig.~\ref{fig6}.

The real part $\Sigma'$ of the computed self-energies for both spin channels and the charge channel are plotted in the last column of Fig.~\ref{fig6} for all three materials. Here, again we see two energy scales of different characteristics. All the spin and charge components of $\Sigma^{\prime}$ are linear in the low-energy region$-$coming from the linear dispersion
of the fluctuation spectrum along $\Gamma\rightarrow(\pi,0)$ and $\Gamma\rightarrow(\pi/2,\pi/2)$, shown in the left two columns$-$giving a total dispersion renormalization of the order of 2-3, consistent with experiments. $\Sigma'$ attains a peak around $200$meV [$400$~meV for electron doped] which means that the resulting electronic dispersion also undergoes a change in its renormalization behavior, yielding the high-energy kink.

The LEK in $\Sigma^{\prime}$ occurs at $70$meV for LSCO and Bi2212 and at $50$meV for NCCO, consistent with ARPES data (black lines). The ARPES-derived $\Sigma'$ usually shows a more pronounced peak at the low-energy kink, rather than a break-in-slope.  This is partially due to the assumed form of the bare dispersion, which is taken as a straight line from the Fermi level to the dressed band at a high energy usually chosen at -300~meV, rather than the correct LDA band. Note that due to the Kramers-Kronig relation, the imaginary part of the self-energy increases gradually with energy and does not show any characteristic features at these two kink energies but shows a maximum at the energy where $\Sigma^{\prime}$ changes sign. Therefore, the spectral weight gradually decreases at these two energy scales.


\begin{thebibliography}{99}
\bibitem{foot_mode_shift}The bosonic mode energy in $\chi^{\prime\prime}({\bm q},\Omega)$ shifts to higher energy in $\Sigma({\bm k},\omega)$ due to gap opening as well as band-structure effects, embedded in the Green's function, given in Eq.~\ref{selfenergy}. Such energy shift becomes more dominant when the VHS in DOS approches the Fermi level with doping. In principle, the energy scale of $\Sigma({\bm k},\omega)$ can also shifts in the single-particle spectra $A({\bm k},\omega)$, because of the same reason.
\bibitem{abfoot1} The doping dependence of the band structure is
  treated within the framework of the rigid band model
  [Ref. \onlinecite{new1}], which is expected to be a good
  approximation for doping away from the CuO$_2$ planes. It will be
  interesting to further examine doping effects by using first
  principles approaches \cite{new2}.
\end{thebibliography}
\end{document}